# Tip-Induced Molecule Anchoring in Ni-Phthalocyanine on Au(111) Substrate

Yong Chan Jeong,[1,†] Sang Yong Song,[1,†] Youngjae Kim,[1] Youngtek Oh,[2] Joongoo Kang,[1,*] and Jungpil Seo[1,*]

[1]*Department of Emerging Materials Science, DGIST, 333 Techno-Jungang-daero, Hyeonpung-Myun, Dalseong-Gun, Daegu 711-873, Korea*
[2]*Samsung Advanced Institute of Technology, Suwon 443-803, Korea*

**ABSTRACT**

Pinning single molecules at desired positions can provide opportunities to fabricate bottom-up designed molecular machines. Using the combined approach of scanning tunneling microscopy and density functional theory, we report on tip-induced anchoring of Ni-phthalocyanine molecules on an Au(111) substrate. We demonstrate that the tip-induced current leads to the dehydrogenation of a benzene-like ligand in the molecule, which subsequently creates chemical bonds between the molecule and the substrate. It is also found that the diffusivity of Ni-phthalocyanine molecules is dramatically reduced when the molecules are anchored on the Au adatoms produced by bias pulsing. The tip-induced molecular anchoring would be readily applicable to other functional molecules that contain similar ligands.

**INTRODUCTION**

Molecules are a unique building block for designing multifunctional nanodevices in a bottom-up fashion[1-11]. Molecules have already been used to realize quantum electronic devices.[1-6] As molecules consist of a few tens of atoms, they inherently exhibit quantum-mechanical natures. Such molecules are widely utilized in developing quantum computers, field-emission transistors, and so on. Molecules are further used as functional, mechanical machines in which chemical or electrical energy is needed to modulate the designed mechanical motions. Examples include electrically driven molecular motors, DNA actuators, and single-molecule propellers.[7-11] Like this, molecules possess two complimentary properties, and the future of molecular devices lies in the ways in which these properties can be systematically controlled. Thus, investigating the interactions between the electric and mechanical properties of a molecule is an important requirement toward tailoring multifunctional molecular devices.

A recent study by Jiang et al. using Fe-phthalocyanine molecules has demonstrated the possibility of such integrated molecular devices in which the molecular mechanical diffusivity was controlled by an external electric field.[12] Furthermore, Comstock et al. and Stock et al. have shown that azobenzene molecules on an Au(111) substrate and Cu-phthalocyanine molecules on a Cu(111) substrate were chemically adsorbed on the substrate by electric pulses.[13,14] Although these studies highlighted the diffusive behaviors of molecules on noble metal substrates, the general mechanism of the adsorption of the molecules onto substrates by electric pulses has not been clearly revealed to date. Here, we report a both experimental and theoretical study combining scanning tunneling microscopy (STM) and density-functional theory (DFT) to reveal the process of the tip-induced anchoring of Ni-phthalocyanine molecules (NiPCs) on an Au(111) substrate. We have demonstrated that the dehydrogenation of the benzene-like ligands in the NiPCs plays an important role in the mechanical anchoring of these molecules. Furthermore, we found that the molecules readily anchored on the Au adatoms of the substrate. Thus, the diffusivity was dramatically modified, and the motion was confined within the herringbone reconstructions of the substrate.

**METHODS**

The experiment has been performed using a home-built variable temperature STM with the base vacuum pressure $p < 8.0 \times 10^{-11}$ torr. We used an electrochemically sharpened tungsten tip to acquire topographic images and spectroscopic data of NiPCs on the Au(111) substrate. The (111) surface of Au is particularly chosen due to its chemical stability and characteristic surface reconstruction. The substrate was cleaned by the repeated cycles of Ar sputtering in the pressure of $1.5 \times 10^{-5}$ torr for 10 min and annealing at 600º C for 15 min. NiPCs (Sigma Aldrich Ni(II) Phthalocyanine, P/N: 039453.06) were thermally evaporated onto the substrate using a home-made Knudsen cell. The density of the molecules deposited on the substrate is about 10 NiPCs per 100 $nm^2$. All the measurements were done at the temperature of 77 K except if otherwise indicated. The differential conductivity spectra were obtained by a standard lockin technique of the modulation frequency $f = 718$ Hz and 10 mV root-mean-square amplitude.

The calculations were done using the generalized gradient approximation (GGA-PBE[15]) to DFT, as implemented in the VASP code.[16,17] Plane waves with the kinetic energy cutoff of 400 eV and projector-augmented wave potentials[18] were used for our DFT simulations. We used a 6-layer Au(111) slab in a (6 × 6) supercell for the substrate and a (6 × 6) k-point mesh for Brillouin-zone sampling. For the calculations of the chemisorption of a NiPC pair, a (12 × 6) supercell of Au(111) was used. Atomic structures were relaxed within 0.05 eV/Å.

**EXPERIMENTAL RESULTS**

A NiPC is structurally based on a porphyrin, and four benzene-like ligands are attached to the sides of the base molecule. Thus, such a molecule contains four-fold rotational symmetry and mirror symmetries (point group $D_{4h}$) [Fig. 1(h)]. At a temperature of 8 K, the NiPCs on Au(111) were separately visible and mostly stationary (see Fig. S1 in the Supporting Information). The NiPCs, which were ordered by finite distances, indicated there were strong repulsive interactions and weak attractive interactions between the molecules.[19-21] The attractive interaction is most likely the van der Waals force. The repulsive interaction is associated with the substrate-mediated charge transfer process,[19] and it is responsible for the rigid scattering among the molecules at higher temperatures (see Fig. S2 in the Supporting Information).

In contrast to the scan image at 8 K (see Fig. S1 in the Supporting Information), the NiPCs were highly diffused on the Au(111) substrate at 77 K (Fig. 1). The image acquired with a +1.0 V sample bias depicted inversed contrast between the herringbone ridges and valleys [Fig. 1(a)]. At first glance, the molecules were effectively confined within the herringbone ridges. From careful study, however, we determined that the NiPCs diffused freely across the herringbone structures (see Fig. S3 in the Supporting Information). The image measured with a -1.0 V sample bias displayed well-ordered Au herringbone structures as if there were no NiPCs on the surface [Fig. 1(b)]. This bias dependence is due to the vertical dipole moment of phthalocyanine on the metal substrate in which a positive bias decreases the potential energy for the molecules to gather the NiPCs beneath the tip and a negative bias repel the NiPCs from the tip.[12,22]

Interestingly, we have found that the molecules cease diffusing and are tightly anchored on the substrate when tunneling electrons are injected through the molecules by bias pulsing, as shown in Figs. 1(a)-1(d). In Fig. 1(a), the tunneling junction between the tip and substrate was stabilized with the bias voltage ($V_{bias}$) of 1 V and the set-point current (I) of 100 pA. In this condition, the molecules were concentrated in the tip below. Subsequently, bias pulses of 4 V and 10 ms were applied to the system at the positions of the marks in Fig. 1(b). During the pulsing the feedback was closed. Remarkably, the molecules were pinned at the pulsed positions and remained immobile during the scans even with negative bias [Figs. 1(c) and 1(d)].

To investigate the effect of the bias pulse on the system, we monitored the variation of the tunneling current during the application of the bias pulse. We found that the tunneling current sharply increased upon the application of the pulse [see Fig. S4 in the Supporting Information]. Therefore, the bias pulse should be understood as a current pulse that flows through the molecules. When we measured the current pulse as a function of the pulse voltage, they showed an exponential-like relationship [see Fig. S4 in the Supporting Information]. Using the relationship, we have plotted the number of the anchored molecule as a function of the pulse power, which has shown the strong dependence between them [Fig. 1(g)]. This signifies that the injected tunneling current is a crucial factor in the mechanism of the molecule anchoring rather than the electric field effect. Furthermore, we designed the experiment in which the electric field was applied between the tip and the sample without flowing the tunneling current [see Fig. S5 in the Supporting Information]. We found that the molecules did not anchor on the surface when the current was absent. This strongly supports that the tunneling electrons rather than the electric field effect cause the molecule anchoring.

We found that the four-fold rotational symmetry was broken in the anchored NiPCs and that the point group was reduced to $C_{2v}$. The molecules in Fig. 1(d), imaged so as to depict the ligand rings, were partially damaged. The zoomed-in image of a typical anchored NiPC is shown in Fig. 1(e), and its height profiles are depicted in Fig. 1(f). The damaged ligand was about 20% shorter than the intact ligands. We attribute this height shrinkage to the tunneling-electron-induced dehydrogenation of the ligand ring.[23-25] To contrast the electronic states between the damaged ligand and the intact ligands, we performed scanning-tunneling-spectroscopy (STS) measurements on the individual ligands (Fig. 2). Fig. 2(a) shows the zoomed-in image of the anchored molecules. The ligand denoted as 1 was damaged, while the others were intact. The measured local densities of states (LDOS) of both the damaged and intact ligands were similar in the filled states. In contrast, there was a significant difference in the empty states in which the LDOS of the damaged ligand was dramatically enhanced [Fig. 2(c)]. In order to confirm that the damage resulted from the bias pulse, we positioned the STM tip over Ligand 4 in Fig. 2(a). By applying the same pulse to the ligand, we observed that the height of the ligand instantly shrank [Fig. 2(b)] and that the spectrum then conformed to that of the damaged one [Fig. 2(d)]. Therefore, conclusively, the ligand damage was closely associated with the bias pulse, and the molecular anchoring resulted from the ligand damage.

**CALCULATIONS AND DISCUSSION**

To understand the tip-induced molecular anchoring at the atomic scale, we examined the energetics of various adsorption states of the NiPCs on the Au(111) surface using DFT calculations. Atomic structures of various chemisorption states of a NiPC on the Au(111) substrate are identified in Fig. 3. In the absence of any defects, a NiPC is weakly physisorbed on the Au(111) surface. Therefore, the chemisorption of the NiPC on the Au surface requires the formation of a tip-induced defect either in the molecule itself or on the substrate surface. We first considered the dehydrogenation of a ligand of the NiPC for possible, tip-induced damage in the molecule. For an isolated NiPC on the substrate, we found that the dehydrogenation of the ligand should occur as the result of a pair of reactions, producing a $H_2$ molecule [Fig. 3(a), denoted by 4]. When two C-H bonds of the dehydrogenated ligand are broken, the two remaining C atoms of the ligand are then available to form two new C-Au bonds with the Au atoms on the surface [Fig. 3(a)]. The C-Au bond length ($d_{C-Au}$) is 2.08 Å. The 90° rotational symmetries of the NiPCs are largely broken by the chemisorption-induced bending of the NiPC [Fig. 3(e)], findings that agree with the experiment results.

The DFT calculations show that the tip-induced chemical adsorption of the NiPCs in Fig. 3(a) is an endothermic reaction. Here, we define the formation energy ($E_{form}$) of the chemisorbed NiPCs as the energy of the chemisorption state with respect to that of the physisorption state [Fig. 4]. The $E_{form}$ of the NiPC in Fig. 3(a) is calculated to be 3.0 eV. The large dehydrogenation energy, which is associated with the strong C-H bonds of the ligand ring, is responsible for the large $E_{form}$. Therefore, the chemisorpition of a NiPC can occur when enough energy is transferred from the tunneling electrons to the molecule. When the bias pulse is applied, the injected electrons will initially occupy the empty states of the physisorbed NiPC. The excitation energy of the injected electrons in the NiPC can be efficiently dissipated to the kinetic energy of the H atoms in the NiPC through electron-phonon coupling.[26] Consequently, the H atoms can be desorbed from the NiPC, forming an $H_2$ molecule. Then, the dehydrogenated benzene-like ligand bends toward the substrate to form two C-Au bonds [Fig. 3(e)]. The formed $H_2$ molecule diffuses out from the anchored NiPC, and it is no longer available for the reverse exothermic chemical reaction. The energy cost for the second anchoring of the NiPC via the dehydrogenation of the ligand denoted by 3 in Fig. 3(a) is 3.1 eV, which is nearly the same as that for the first anchoring [Fig. 4]. The atomic structure of the NiPC with the two damaged ligand rings is shown in Figs. 3(b) and 3(f).

The STS results in Figs. 2(c) and 2(d) showed that the differential conductivity at the energies ranging from 0.6 eV to 1.0 eV was substantially enhanced on the dehydrogenated ligands upon the anchoring of a NiPC. Using DFT calculations, we found that the Au-C bonds of the chemisorption state induced the electronic states coupled to the Au metallic state within the corresponding energy window for which the electron density on the molecule's side was distributed nearly exclusively on the dehydrogenated ligand but not on the remaining intact ligands [Figs. 3(e) and 3(f)]. Therefore, the anchoring-induced electronic states contribute to the enhanced LDOS for the tip positions on the damaged ligand rings, leading to the enhanced differential conductivity in the STS measurements.

In addition to the chemisorption of the dehydrogenated NiPCs, we also explored the possibility that the tip-induced electron injection might have created an Au adatom ($Au_{ad}$), subsequently providing a new binding site for a NiPC via a N-$Au_{ad}$ bond [Fig. 3(c)]. The N-$Au_{ad}$ bond length is 2.29 Å. However, the intrinsic symmetry of a NiPC under 90° rotations is nearly intact after the adatom-mediated adsorption. Furthermore, we found that the energy barrier for the rotational motion of a NiPC around an $Au_{ad}$ is negligibly small. Thus, if it were adsorbed on the $Au_{ad}$, the NiPC would exhibit rotational motions at 77 K, as demonstrated by previous experimental work.[27] Despite its relatively small formation energy ($E_{form}$ = 0.9 eV), rotational molecules were not identified in Fig. 1.

To confirm the adatom-mediated NiPC chemisorption, we applied a pulse with enough of a low-impedance tunnel junction to make a mechanical indentation in the surface via the tip. The pulse was 4 V and lasted 10 ms, and the impedance of the tunnel junction was set to half of that utilized in Fig. 1 ($V_{bias}$ = 1 V, I = 200 pA). In the scan of the pulsed area, we observed the protrusions associated with the indentation and several frozen NiPCs near the protrusions. Remarkably, we found that the majority of the imaged molecules did not have damaged ligands and thus preserved their $D_{4h}$ symmetries [Fig. 5(a) and 5(b)]. Only one molecule was identified as a dehydrogenized NiPC, which is marked with an arrow and the number 1 in Fig. 5(b). We attributed the molecules with the $D_{4h}$ symmetries to the NiPCs anchored on the Au adatoms created by the mechanical indentation. In fact, Hla et al. have shown that Ag adatoms were readily created via tip indentation on a Ag substrate.[28] Similarly, it is highly likely that the Au adatoms were created by the tip indentation and that the NiPCs were anchored on the adatoms, as the DFT calculations suggested.

By successive scans of the area, we found that the dehydrogenated NiPCs were immobile, results that were consistent with those depicted in Fig. 1. In comparison, the locations and orientations of the NiPCs with $D_{4h}$ symmetries continuously adjusted to the perturbations by the molecular scattering from the background [Fig. 5(c)]. By repeating the scan, the molecules eventually diffused away from their original positions due to the influence of the scanning tip. The diffusivity was dramatically decreased, and the motion was restricted within the herringbone valleys of the Au substrate [Fig. 5(d)]. Although the NiPC anchored on an $Au_{ad}$ was more energetically stable than a dehydrogenated NiPC, the voltage pulse without the mechanical indentations barely created any surface adatoms. This pulse was most likely the excitation energy of the injected electrons occupying the high-energy states of the Au substrate and was quickly dissipated through the electron–electron scattering in the metallic states of the substrate.[27]

We focused on the interactions between a single NiPC and the substrate. Interestingly, however, Fig. 1(d) shows that the tip-induced electron injection created not only isolated NiPCs but also NiPC pairs anchored on the substrate. The formation of NiPC pairs was unexpected, considering the repulsive intermolecular interaction at such a short distance. The net formation energy of two well-separated, anchored NiPCs was twice as large as the formation energy of a single NiPC (i.e., $E_{form}$ = 6.0 eV/pair). Therefore, two NiPCs that were just joined at a close distance had even larger formation energies. To explain the tendency of the NiPC pairing through the tip-induced chemical reaction, we propose atomic structures for a NiPC pair in which each benzene-like ligand of the NiPC pair donates a single H atom to form a $H_2$ molecule [Fig. 3(d)]. Note that the proposed NiPC pair involves only two broken C-H bonds. Hence, the formation energy of the NiPC pair is as low as 3.5 eV, which is compatible with that of a single NiPC [Fig. 4]. Thus, if two NiPCs happened to come sufficiently close before the tip-induced chemical reaction occurred, a NiPC pair could be formed at the energy cost of the single NiPC chemisorption.

**CONCLUSION**

In summary, we have found that the NiPCs are anchored on Au(111) substrate when bias pulses are applied. From the combined study between the STM and DFT calculation, it has been shown that the dehydrogenation in the ligands of NiPCs plays an important role in the molecule anchoring. When the molecules are dehydrogenated, the C-Au bonds are created between the molecules and the substrate. In addition, we have shown NiPCs are anchored on the Au adatom by forming the N-Au bonds between the molecules and adatoms. We have demonstrated the adatoms are created by the mechanical indentation induced by the bias pulse. By showing the clear correlation in the yield-vs-power plot, our work establishes that when bias pulses are applied, the tunneling current is responsible for the anchoring of molecules. We also rule out the electric field effect as an alternative cause of the molecule anchoring. We believe that our quantitative analysis provides useful insights into understanding tip-induced chemical reactions and will stimulate further experimental and theoretical research exploring the energy transfer process of injected electrons in molecules.

ACKNOWLEDGMENT

This work was supported by the Leading Foreign Research Institute Recruitment Program (2012K1A4A3053565) and DGIST R&D Program (15-BD-0403, 14-HRMA-01) through the National Research Foundation of Korea (NRF) funded by the Ministry of Science, ICT and Technology of Korea.

SUPPORTING INFORMATAION

NiPCs on Au(111) Measured at 8 K; Molecule Fence Made of NiPCs; NiPCs Diffuse Across the Herringbone Ridges of Au Surface; The Anchored NiPCs Depending on the Pulse Power; Electric Field Effect in the Molecule Anchoring

AUTHOR INFORMATAION

*Corresponding authors: joongoo.kang@dgist.ac.kr, jseo@dgist.ac.kr

†These authors contributed equally to this work.

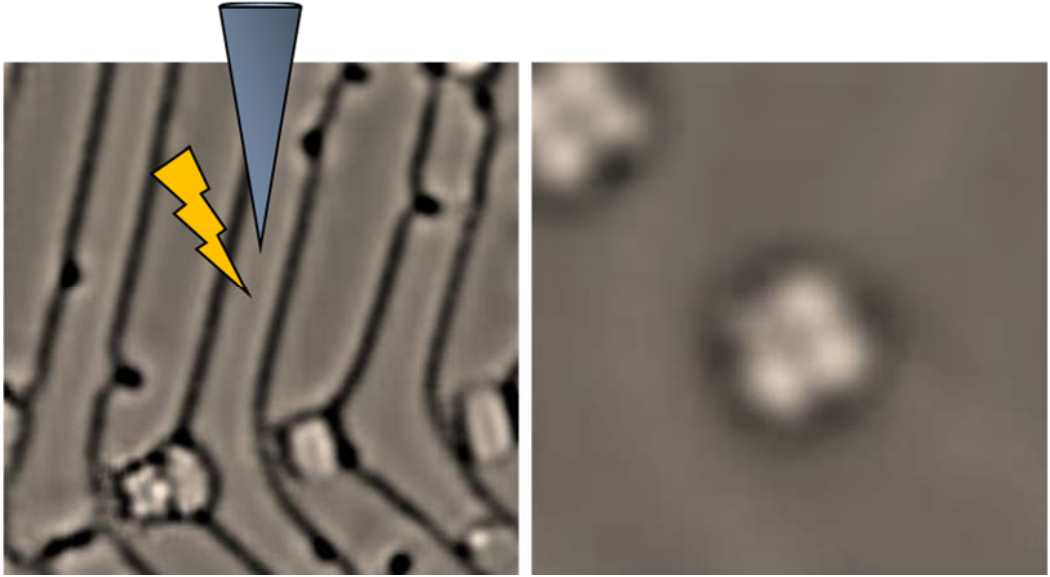

**Table of Contents Graphic**

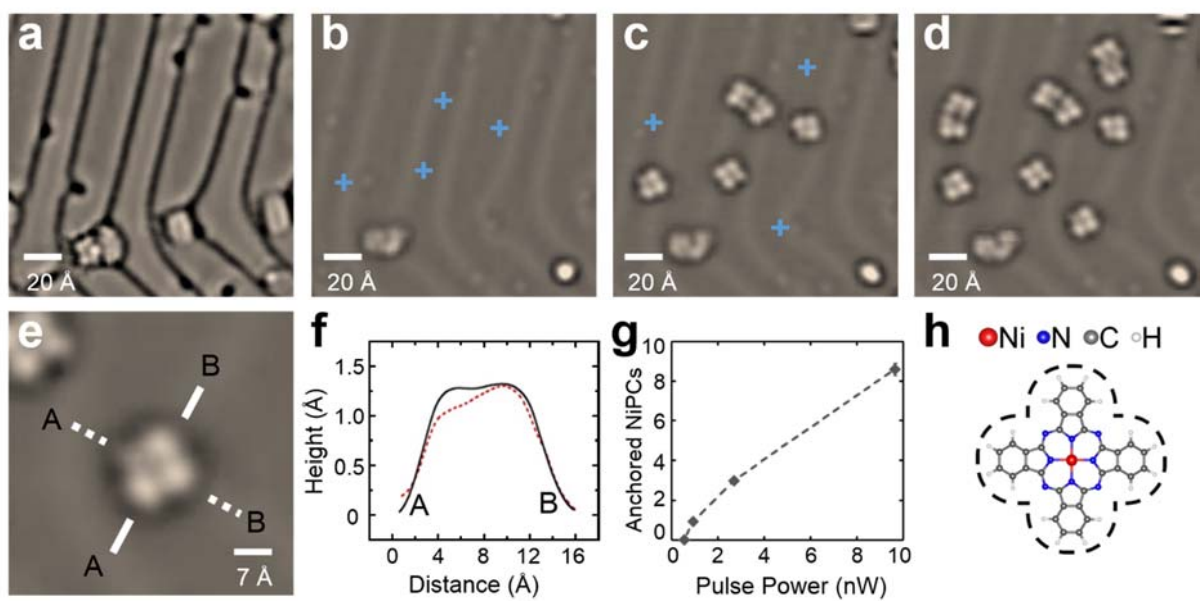

**Figure 1.** Molecular Anchoring Depending on Electric Pulses. (a) With $V_{bias}$ = 1 V, the topographic image of the NiPCs shows inversed contrast between the herringbone ridges and valleys. (I = 100 pA) (b) The area is the same as that shown in (a) but scanned with a $V_{bias}$ = -1 V (I = 100 pA) (c) The bias voltage was switched to 1 V. The bias pulses of 4 V and 10 ms with feedback were applied at the position of the blue cross marks shown in (b). NiPCs were pinned at the locations of the pulses. The image is scanned with $V_{bias}$ = -1 V and I = 100 pA. (d) Further pinned molecules were imaged at the locations of the pulses marked in (c). The conditions for pulsing and scanning are same with (c). (e) Zoomed-in image of an anchored NiPC. (f) The height profiles of the anchored NiPC. (g) The number of the anchored molecules as a function of the pulse power. (h) The schematic of the NiPC that contains four ligand rings.

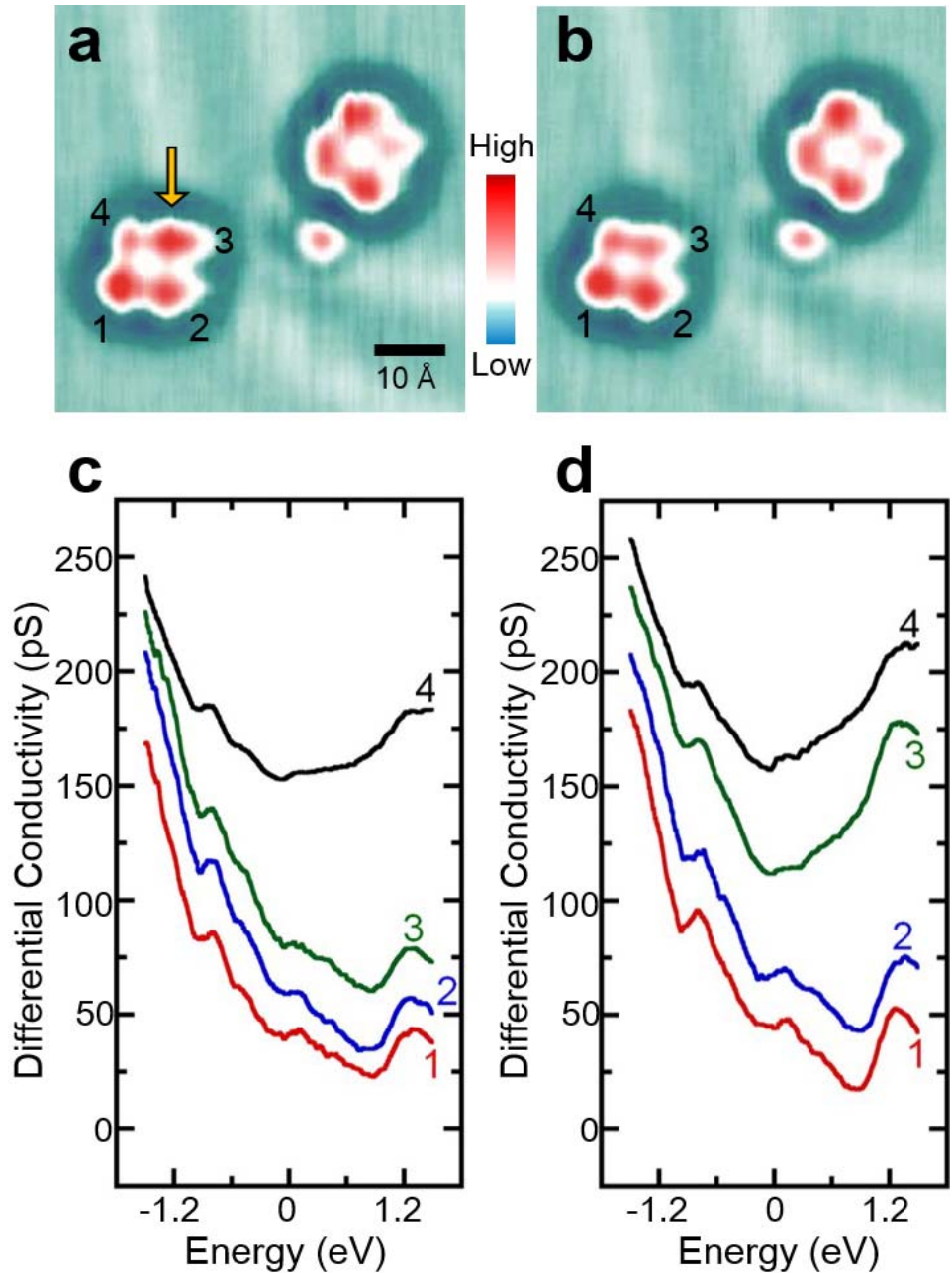

**Figure 2.** STS Measurements of a NiPC. (a) The ligands in the NiPC are identified with numbers. Ligands 1, 2, and 3 are apparently higher than Ligand 4. The image was scanned with $V_{bias}$ = -1 V and I = 100 pA. (b) The bias voltage was switched to 1 V. Bias-pulsing of 4 V and 10 ms was applied at Ligand 3, marked with a yellow arrow in (a). The image shows that Ligand 3 shrank immediately after the pulse. The scanning condition was same with (a). (c) Differential conductance spectra measured on the individual ligand rings ($V_{bias}$ = 1.5 V, I = 150 pA). Ligand 4 shows an enhanced DOS in the empty states, as compared to Ligands 1, 2, and 3. (d) After pulsing, the spectrum of Ligand 3 corresponded to that of Ligand 4. The spectra were offset vertically for clarity in (c) and (d).

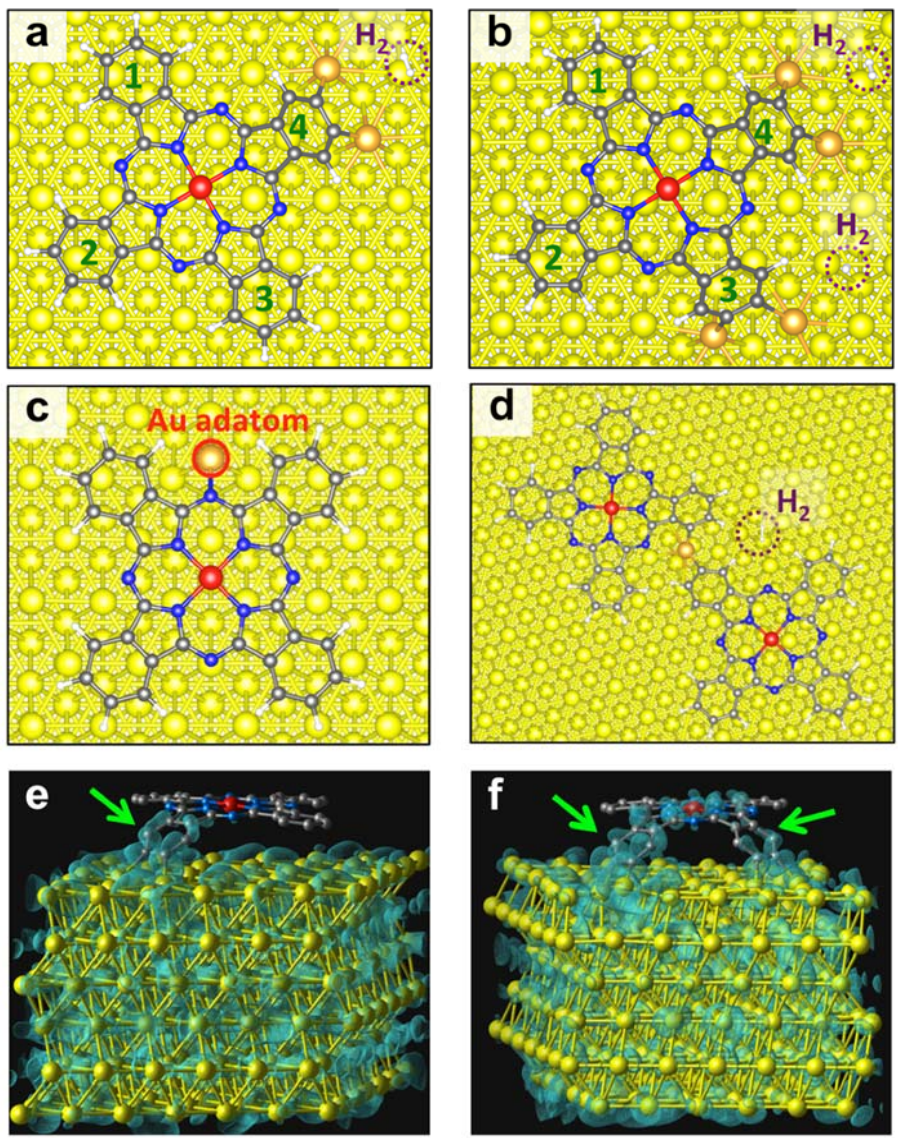

**Figure 3.** Atomic Structures of Various Chemisorption States of a NiPC on a Defect-Free Au(111) Substrate. (a) Chemisorption of NiPC with a single dehydrogenated ligand (denoted by 4) bonding to two surface Au atoms (colored orange). The desorbed H atoms formed an $H_2$ molecule, as marked by a circle. (b) The tip-induced anchoring of a second, dehydrogenated ligand at the position denoted by 3. (c) Au adatom-mediated chemisorption of a NiPC. (d) Atomic structure of a NiPC pair. The pairing of NiPCs involves the formation of a single $H_2$ molecule (marked by a circle). (e)-(f) Charge density plots of the chemisorption-induced electronic states at 1 eV for the NiPC in Figs. 3(a) and 3(b), respectively. The green arrows indicate the electronic states of the dehydrogenated ligand which are coupled to the Au metallic states.

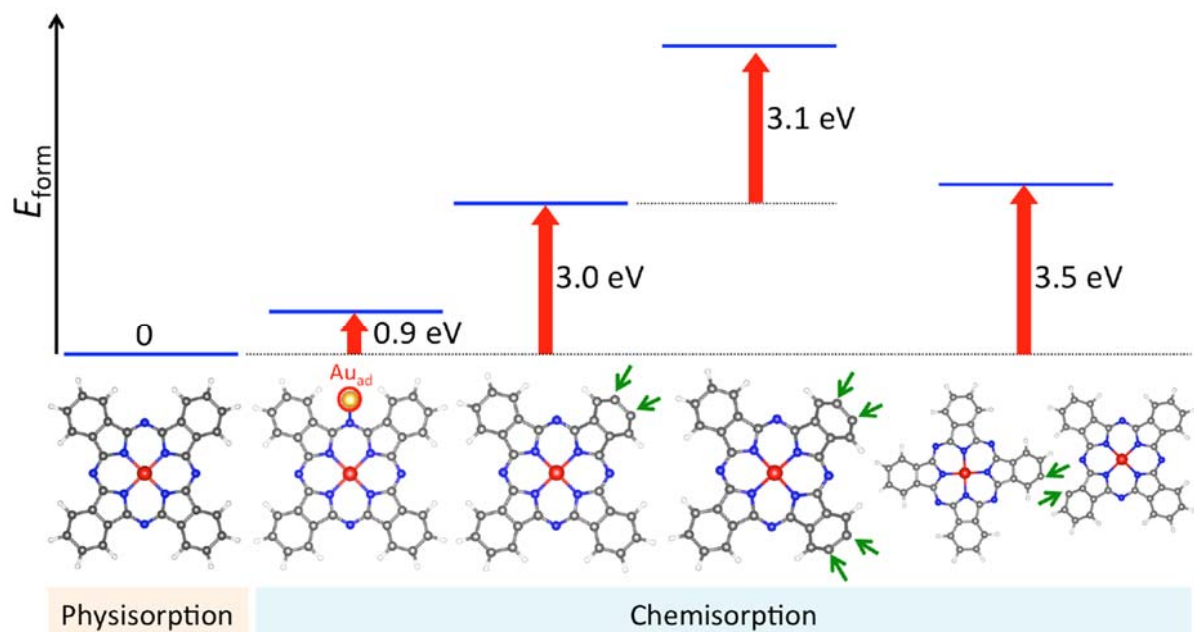

**Figure 4.** Comparison of the Formation Energies ($E_{form}$) of the Adsorbed NiPCs. The $E_{form}$ of the physisorbed molecule (left) is referenced to zero. The positions of the dehydrogenated C atoms of the NiPC are marked by arrows.

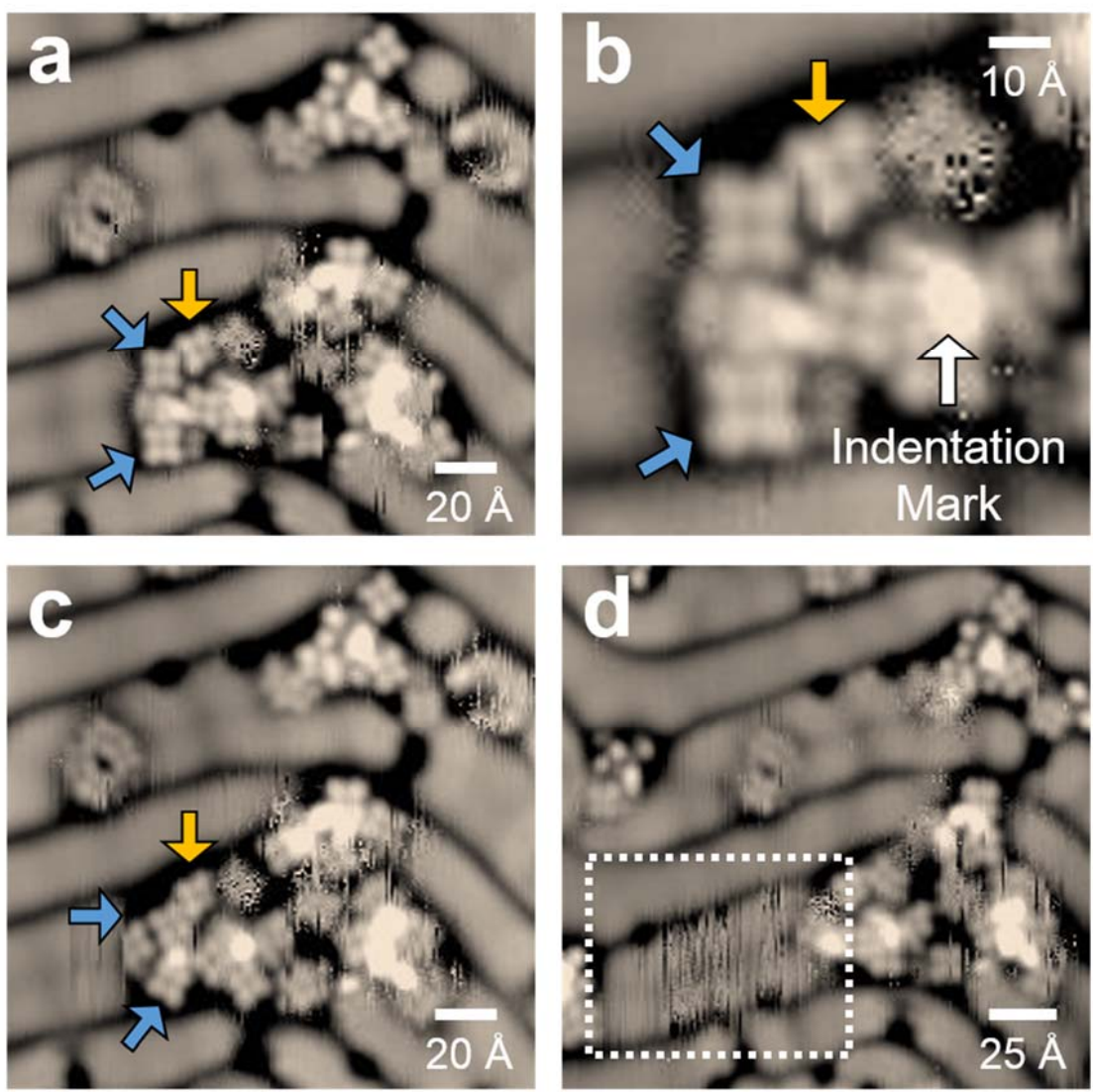

**Figure 5.** NiPCs Anchored on Au Adatoms. (a) Image of NiPCs after the surface indentation induced by a bias pulse. (b) A typical dehydrogenated NiPC is marked with a yellow arrow. The NiPCs that preserve $D_{4h}$ symmetries were found anchored on the Au adatoms. Some of them are marked with blue arrows. (c) The dehydrogenated NiPC was immobile during the scans. In contrast, the NiPCs anchored on the adatoms continuously changed their locations and orientations. (d) By repeating the scan, the NiPCs anchored on the adatoms eventually diffused out from their original positions (dotted box). The images are scanned with $V_{bias} = 1$ V and I = 100 pA.

# Supporting Information

# Tip-Induced Molecule Anchoring in Ni-Phthalocyanine on Au(111) Substrate

Yong Chan Jeong,[1,†] Sang Yong Song,[1,†] Youngjae Kim,[1] Youngtek Oh,[2] Joongoo Kang,[1,*] and Jungpil Seo[1,*]

[1]*Department of Emerging Materials Science, DGIST, 333 Techno-Jungang-daero, Hyeonpung-Myun, Dalseong-Gun, Daegu 711-873, Korea*

[2]*Samsung Advanced Institute of Technology, Suwon 443-803, Korea*

*Corresponding authors: joongoo.kang@dgist.ac.kr, jseo@dgist.ac.kr

†These authors contributed equally to this work.

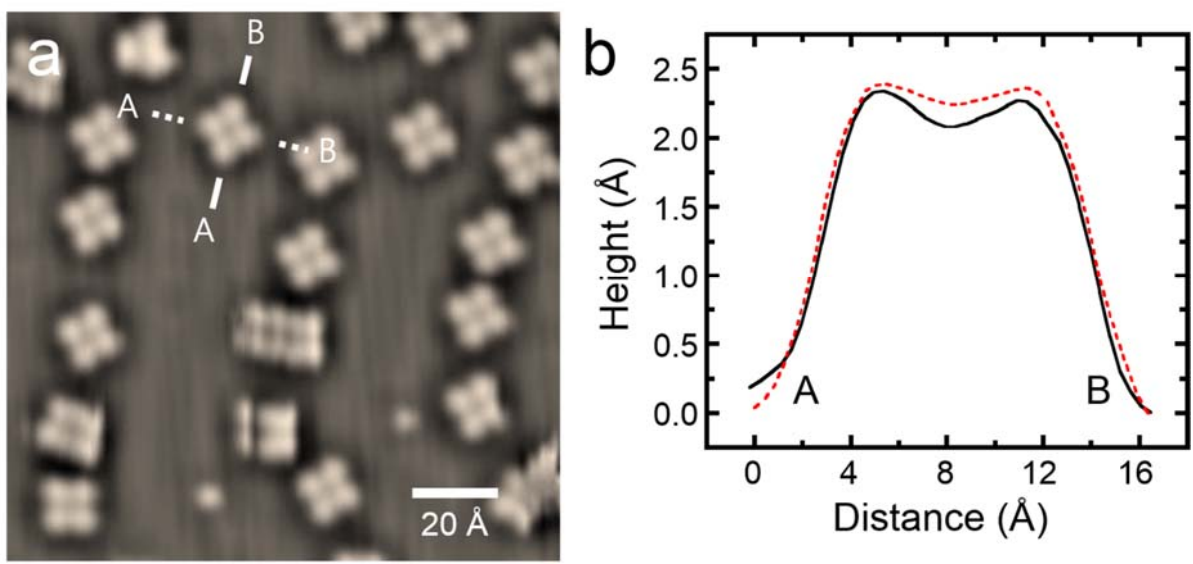

**Figure S1. NiPCs on Au(111) Measured at 8 K. (a)** The NiPCs are stationary and well-separated at 8 K. The NiPCs possess four-fold rotational symmetry and mirror symmetries ($D_{4h}$). **(b)** The height profiles of a typical NiPC in (a).

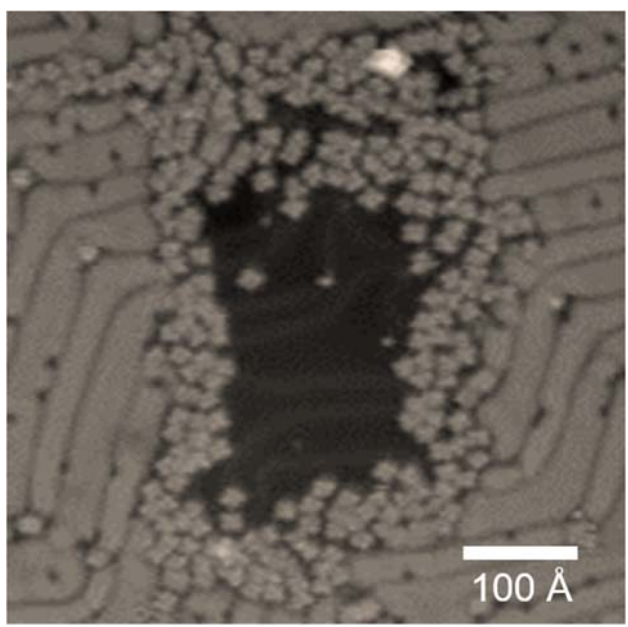

**Figure S2. Molecule Fence Made of NiPCs.** NiPC molecules are anchored through the tip-induced dehydrogenation. Once the anchored molecules form closed path and the diffusive molecules inside the closed path are further anchored by the tip, the molecules cannot move in through the closed path. This observation indicates the NiPCs act as rigid bodies in the scattering process.

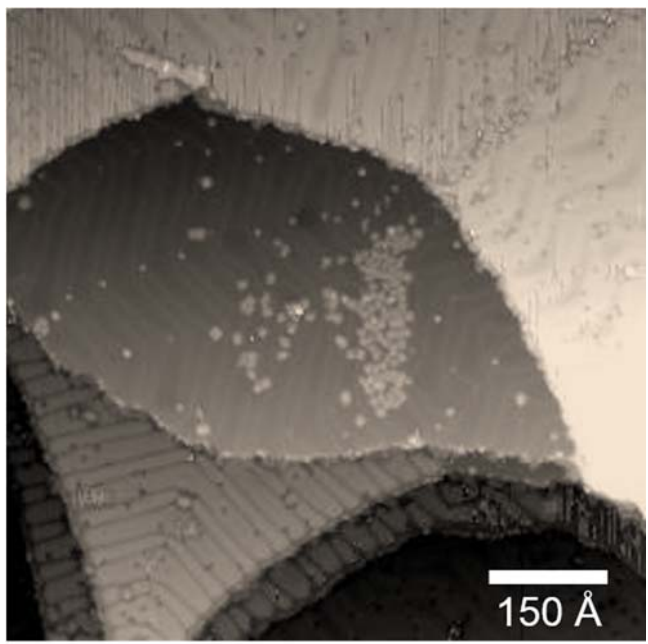

**Figure S3. NiPCs Diffuse Across the Herringbone Ridges of Au Surface.** All NiPCs are anchored in a terrace by the tip-induced dehydrogenation. Although the anchored NiPCs are localized, the depletion of the molecules is global in the terrace. This indicates the NiPCs diffuse freely across the herringbone ridges of Au surface.

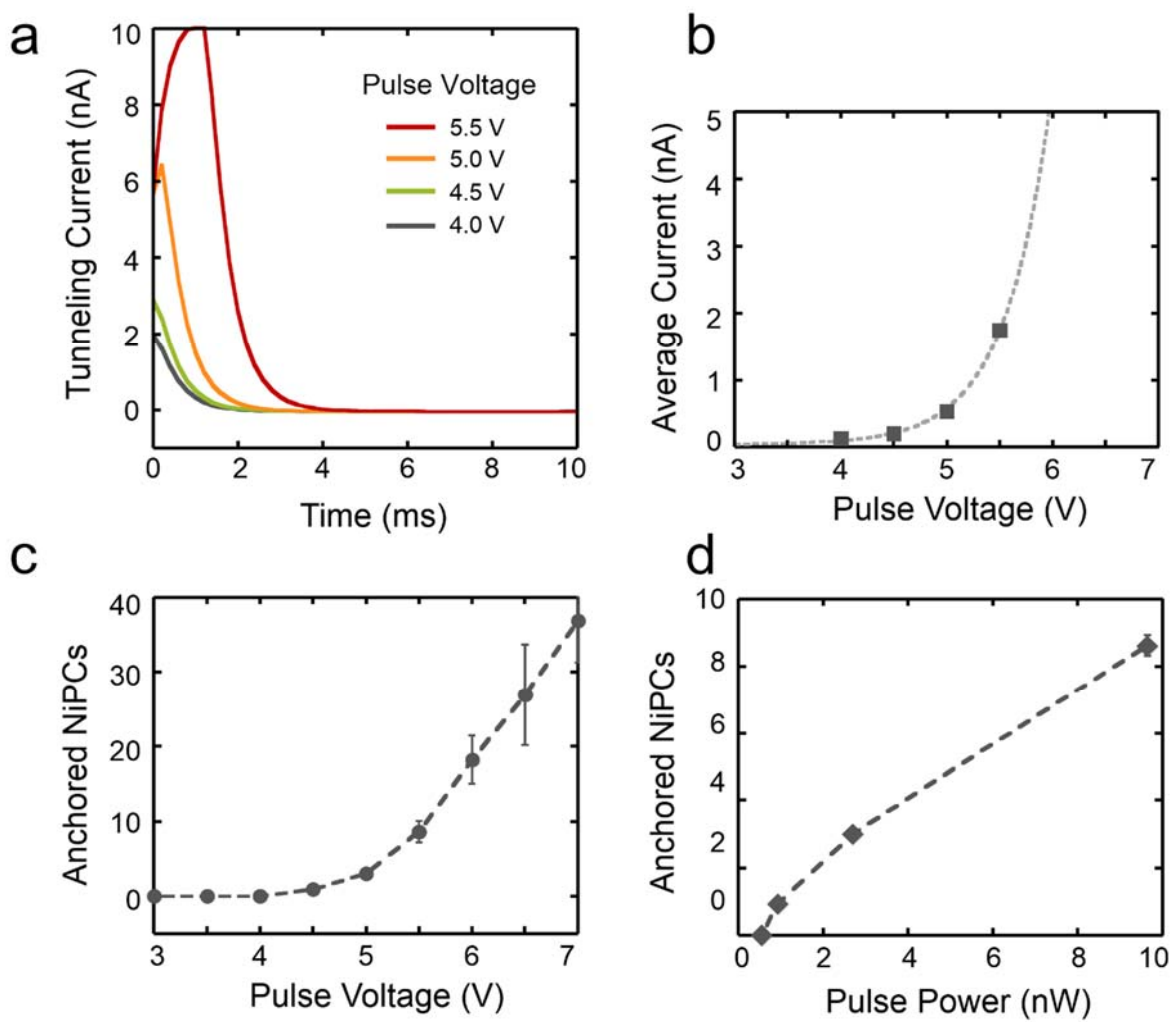

**Figure S4. The Anchored NiPCs Depending on the Pulse Power.** (a) We have recorded the tunneling current during the application of the bias pulses. We have found the current induced by the pulse is increased non-linearly with the pulse voltage. (b) We have calculated the average current during the pulse time of 10 ms from (a). The graph shows that the average current non-linearly depends on the pulse voltage. The dotted line is an exponential fitting. (c) The anchored molecules as a function of the pulse voltage. (d) Using the results from (b) and (c), we have plotted the anchored molecules as a function of the pulse power.

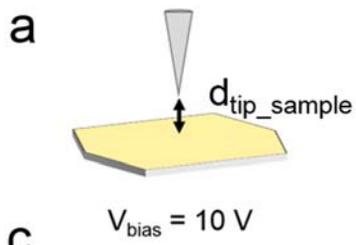
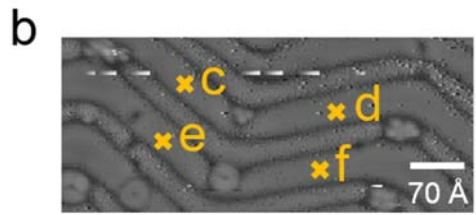
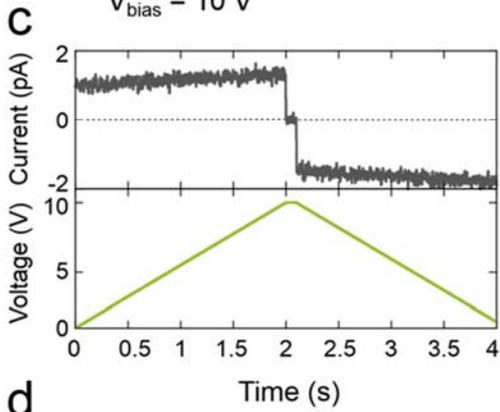
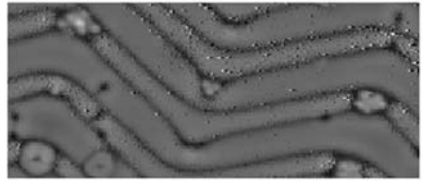
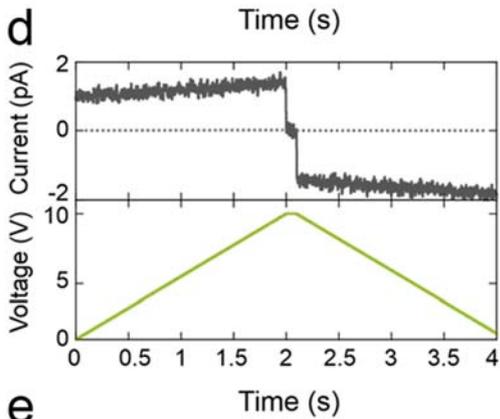
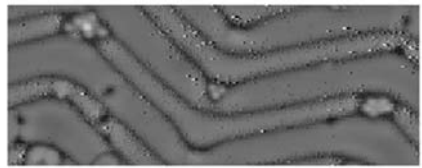
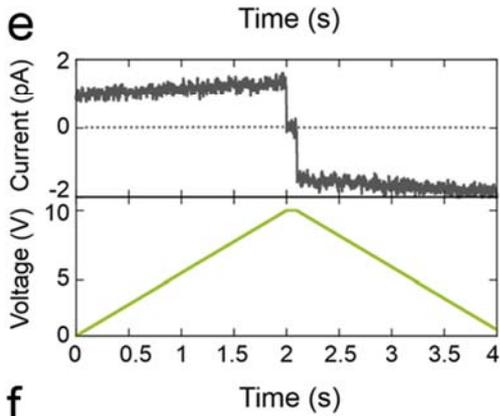
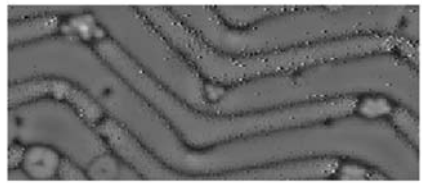
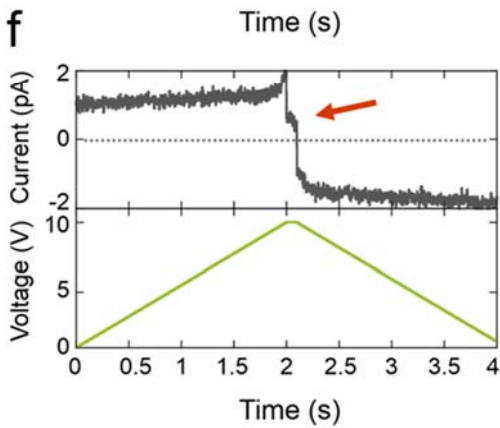
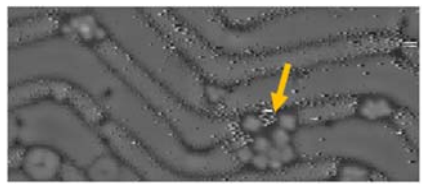

**Figure S5. Electric Field Effect in the Molecule Anchoring** (a) We have measured the distance between the tip and the sample by moving the tip to the sample until they touch with the preamp gain of 1. When the bias voltage was 1 V and the set-point current was 100 pA, the distance was found as 20 Å. When we apply the bias pulse of the 4 V, the magnitude of the electric field is calculated by about 4 V / 20 Å (= 0.2 V/Å) by supposing the two parallel electrodes. The purpose of the experiment is applying the electric field in the STM junction without current flows to investigate the role of the electric field in the molecule anchoring. For this purpose, we will apply higher electric field than 0.2 V/Å without the current. (b) The image shows the area before applying the electric field. Each mark indicates the spot we applied the electric field. The detail of the experiment is in the labelled section. (c) We parked the tip with the bias voltage of 1 V and set-point current of 100 pA. Then, we moved the tip 20 Å backward so the distance between the tip and the sample became 40 Å. The voltage is ramped from 1 V to 10 V for 2 seconds, and stayed 20 ms with 10 V, and ramped down. No current is found during the application of 10 V. Note that the displacement current during the ramping is due to the capacitance coupling between the signal wires of STM. The electric field is ~ 10 V / 40 Å (= 0.25 V/Å), which is slightly larger than the 0.2 V/Å. Nevertheless, we could not find any anchored molecule on the surface. (d), (e) We continued to decrease the distance between the tip and the sample in the same experimental conditions. Neither the current flows nor the anchored molecules were observed until the distance of 16 Å. (f) When the distance became 14 Å, the current was detected in the junction as marked with the arrow. When the current flew, we found the molecules were anchored on the surface (marked by the arrow in the image). The experiment explicitly shows that the molecule anchoring is due to the reaction of currents rather than the electric field effect.